\documentclass[usenatbib]{mn2e}

\usepackage{graphicx}
\usepackage{epsfig}
\usepackage{float}
\usepackage{amsmath}

\title[Millihertz Optical QPOs in 4U~0614+091]{Millihertz
  Quasi-periodic Optical Oscillations in 4U~0614+091}

\author[Zhang, Hynes \& Robinson]{Y. Zhang$^{1}$\thanks{E-mail:
yzhang@phys.lsu.edu}, R. I. Hynes$^{1}$\thanks{E-mail: rih@phys.lsu.edu} and E. L. Robinson$^{2}$\thanks{E-mail: elr@astro.as.utexas.edu}\\
$^{1}$Dept. of Physics and Astronomy, Louisiana State University,
Baton Rouge, LA 70803, USA \\
$^{2}$Department of Astronomy, University
of Texas at Austin, Austin, TX 78712, USA}

\begin{document}

\date{Accepted . Received ; in original form }

\pagerange{\pageref{firstpage}--\pageref{lastpage}} \pubyear{2011}

\maketitle

\label{firstpage}

\begin{abstract}
  We report the discovery of a 1\,mHz optical quasi-periodic
  oscillation (QPO) in the candidate ultracompact low-mass X-ray
  binary 4U~0614+091. The ultra-low frequency QPO has no X-ray
  counterpart in contemporaneous {\it RXTE}/PCA data and is likely a
  signature of structure in the accretion disk. The QPO can be
  reasonably fitted with a single sine wave but with a phase jump part
  way through the observation, indicating that it is not coherent. We also identify a 48\,min modulation,
  approximately consistent with the suggested orbital period of
  \citet{OB05} and \citet{2008PASP..120..848S}. If this is indeed
  orbital, it supports an identification of 4U~0614+091 as an
  ultra-compact source.
\end{abstract}


\section{Introduction}

Low-mass X-ray binaries (LMXBs) can be divided into black hole
candidates and neutron star systems. Those with orbital periods of
hours to days are expected to have hydrogen-rich mass donors.  A few
ultra-compact X-ray binaries (UCXBs) containing hydrogen-deficient and
possibly degenerate donors can evolve to extremely small binary
separations, with orbital periods as short as a few minutes
\citep{1986ApJ...304..231N}.  Although these systems were initially
assumed to be rare, many candidates have now been found
\citep{2006astro.ph..5722N,2007A&A...465..953I}.

The X-ray source 4U~0614+091 was first optically identified by
\citet{1974MNRAS.169...25M} with a faint ($\sim 18$\,mag), blue,
variable star (V1055~Ori) located in the Galactic plane
\citep{1994A&A...281L..17V}.  Several arguments recently concluded
4U~0614+091 is a UCXB hosting a neutron star.  \citet{2001ApJ...560L..59J} found an enhanced
neon to oxygen ratio in 4U~0614+091 similar to that seen in other
UCXBs.  Through sensitive optical spectroscopy,
\citet{2004MNRAS.348L...7N} and \citet{2006MNRAS.370..255N} found
significant carbon and oxygen emission lines, but no hydrogen or
helium. They argued 4U~0614+091 contains a carbon-oxygen accretion
disk.  Recently, \citet{Madej2010} reported the discovery of a broad emission feature at $\sim 0.7$ keV, which is attribute of O~VIII Ly$\alpha$ emission, in the spectra of 4U~0614+09. This feature has been seen so far in two systems \citep{Madej2011} that are receiving oxygen-rich material from the  CO or ONe white dwarf. It is an endorsement of an oxygen-rich mass donor star.  Thermonuclear type I X-ray bursts were observed by OSO-8
\citep{1978MNRAS.182..349S}, and assuming they were Eddington limited,
\citet{1992A&A...262L..15B} deduced a distance to 4U~0614+091 probably
$<3$\,kpc.  More recently \citet{2010A&A...514A..65K} have re-examined
archival bursts from several satellites and derived a distance of
3.2\,kpc.  These authors argue that the bursts require more helium in
the accreted matter than would be consistent with the optical spectra
leaving the nature of the mass donor still in question.

A straightforward resolution of this puzzle has been elusive due to
the difficulty in identifying an orbital period.
\citet{1990MNRAS.247..205M} found a 10-d photometric modulation which
might be caused by precession of an accretion disc. They then scaled
the precession period of Her~X-1 to 4U~0614+091 to infer the orbital
period. If the ratio of the disc radius to the neutron star Roche lobe
radius, $\rho$ is assumed to be $\sim 1$, $P_{\rm orb} \approx 6.3-9.4
\text{ hr}$. If $\rho \sim 0.5$ then $P_{\rm orb} \approx 2.4-3.3 \text{
  hr}$. Either way, this argument suggests a non-UCXB system with a
hydrogen-rich donor.  However, the faintness of its X-ray emission in
spite of a nearby distance suggest 4U~0614+091 has a very low accretion
rate more consistent with an orbital period $< 1$ hr
\citep{2003ApJ...598.1217D}. Recently, \citet{OB05} reported a
tentative 50\,min optical period and \citet{2008PASP..120..848S} also
claimed a strong modulation of 51.3\,mins. A spectroscopic modulation
on a period close to this may also be present
\citep{2006MNRAS.370..255N}.  These recent results support the case
that 4U~0614+091 is a UCXB system.

The orbital period is not the only periodicity expected. Superhumps at
close to the orbital period can occur in short period (and usually
small mass ratio) LMXBs due to the beat between the orbital period and
a slowly precessing accretion disk \citep{2001MNRAS.321..475H}.
Millihertz quasi-periodic oscillations (QPOs) also occur, but are
rarely observed at optical wavelengths.  The exception is 4U~1626--67
which is a LMXB source containing a 7.66\,s X-ray pulsar in an
ultra-compact 42\,min orbit.  130\,mHz ($\sim 7.7 s$) X-ray pulsations
are seen, a 48\,mHz ($\sim 20.8 s$) QPO is produced by an interaction
between the pulsar's magnetosphere and the accretion disk, and a
1\,mHz ($\sim 16.7$\,min) UV QPO were found
\citep{2001ApJ...562..985C}. Only the 1\,mHz QPO was not detected in
simultaneous X-ray observations. \citet{2001ApJ...562..985C} suggested
that the 1\,mHz oscillations are due to warping of the inner accretion
disk.

We present here new optical photometry of 4U~0614+091 revealing a new
mHz QPO and further evidence for a $\sim50$\,min period.  In
Section~\ref{ReductionSection} we describe our data.  The lightcurves
are analysed in Section~\ref{AnalysisSection} and compared with
contemporaneous X-ray observations in Section ~\ref{XraySection}.
Finally in Section~\ref{DiscussionSection} we discuss the implications
of a $\sim50$\,min orbital period and the possible origin of the QPO.

\section{Observations and Data Reduction}
\label{ReductionSection}

We performed fast optical photometry of 4U~0614+091 on three nights in
December 2007 with the 2.1\,m~(82'') Otto Struve Telescope at McDonald
Observatory, Texas (Table~\ref{table:obs_log}).  Approximately 7 hours
of data were obtained on each night.  For convenience we will refer to
the nights as Night 1, Night 2, or Night 6 for the remainder of this
work. We used the Argos CCD Photometer \citep{2005JApA...26..321M}. This
is a CCD camera designed for fast photometry that was originally
designed for pulsating white dwarfs, but has proven ideal for
interacting binaries.  It is possible to obtain exposures as short as
1 second with negligible dead-time between exposures. Hence it is
ideally suited to relatively short period phenomena, such as UCXB
orbital periods and mHz QPOs.

As listed in Table \ref{table:obs_log}, a total of 8132 CCD image
frames were taken with an exposure time of 10\,sec per frame. Timing
was synced to GPS 1\,s ticks and checked against multiple NTP
servers. All images used a broad $BVR$ filter to maximize the source
count rate.

Custom IDL software written for Argos data was used to subtract
dark current and bias, and apply flat field corrections to the data
before performing aperture photometry of 4U~0614+091 and a brighter
comparison star. All results reported here are based on differential
photometry relative to the comparison star, which showed no signs of variability.

\begin{table}
 \centering
 \begin{minipage}{\linewidth}
 \centering
  \caption{Optical observing log for 4U~0614+091.}
  \label{table:obs_log}
  \begin{tabular}{lccc}
  \hline
   Name     &    Observation Date & Duration  & Exp Time
   \\
    &  &  (sec)& (sec)\\
 \hline
 Night 1 & 12/08/2007 & 27330 & 10 \\
 Night 2 & 12/09/2007 & 27450 & 10 \\
 Night 6 & 12/13/2007 & 26540 & 10 \\
\hline
\end{tabular}
\end{minipage}
\end{table}

\section{Optical Analysis and Results}
\label{AnalysisSection}

\subsection{Lightcurves}

\begin{figure}
\begin{center}
\begin{minipage}[!h]{\linewidth}
\centering
 \epsfig{file=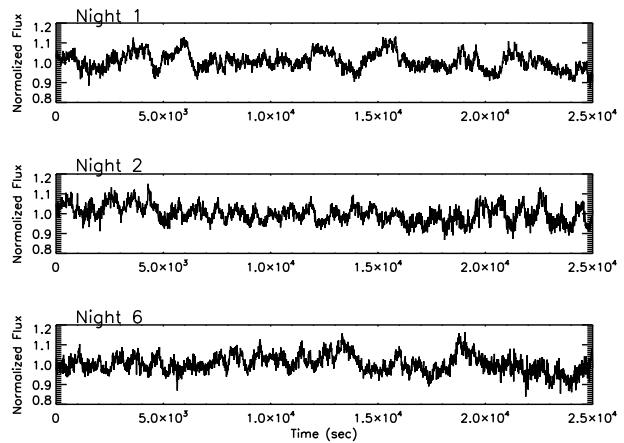,width=\textwidth}

\end{minipage}
\end{center}
\caption{Light curves of 4U~0614+091 during the three nights (top panel: night 1; middle panel: night 2; bottom panel: night 6). All
  fluxes are measured relative to a comparison star and
  normalized to unit mean flux.}
\label{fig:lightcurve}
\end{figure}

Figure~\ref{fig:lightcurve} shows light curves of 4U~0614+091 for our
three observations.  While much of the variability appears to be aperiodic
flickering, an oscillation with a $\sim1000$\,sec period is prominent
throughout night 2 and possibly in shorter segments on
the other two nights.

\subsection{Periodicity Analysis}

\begin{figure}
\begin{center}
\begin{minipage}[!h]{\linewidth}
\centering
 \epsfig{file=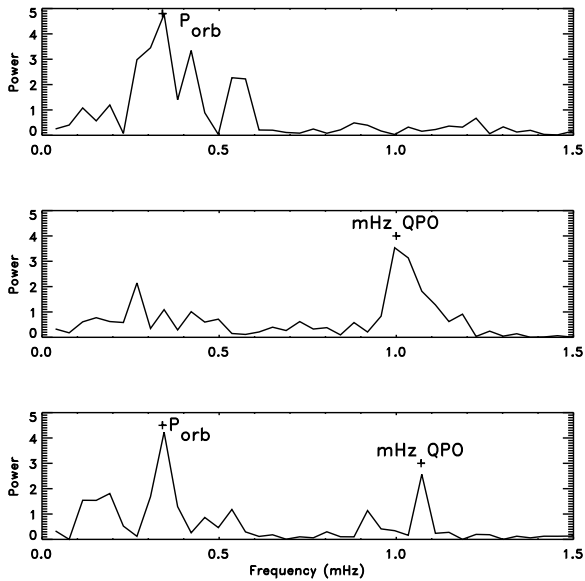,width=\textwidth}

\end{minipage}
\end{center}
\caption{Linearly scaled power spectra of 4U~0614+091 for three
observations (top panel: night 1; middle panel: night 2; bottom panel: night 6).}
\label{fig:linear_po}
\end{figure}

To better quantify the periodicities present,
fast Fourier transforms (FFTs) were computed to derive power spectra.
These power spectra are shown in
Figure~\ref{fig:linear_po}. Two major peaks are recurrent across
multiple nights. A lower frequency at 0.34\,mHz mainly appeared on
Nights 1 and 6, and is coincident with earlier claims of a tentative
$P_{\rm orb}$ \citep{OB05,2008PASP..120..848S}. The higher frequency
signal at 1.03\,mHz is strong only on Night 2, but a weaker signal at
a similar frequency is also detectable on the last night. The two
peaks were fitted with Gaussians in order to measure
characteristic frequencies. Fitting parameters are recorded in Table
\ref{table:obs_fit}. The 1.03\,mHz peak corresponds to a
$16.2\pm0.1$\,min period and the 0.34 mHz peak to a $49.0\pm0.1$\,min
period.

\begin{table}
\centering
 \begin{minipage}{\linewidth}
 \centering
  \caption{Results of fitting Gaussians to linearly scaled power spectra of 4U~0614+091.}
   \centering
  \label{table:obs_fit}
  \begin{tabular}{@{}lllll@{}}
  \hline
   Name    & &   Peak Freq & FHWM & Period
   \\
   & & (mHz) &  (mHz) & (min)\\
 \hline
 Night 1 &Peak 1 & $0.34 \pm 0.01 $& $0.09\rm{ }^{+0.01}_{-0.02} $& $48 \pm 2 $\\
 \\
 Night 2 &Peak 2& $1.03 \pm 0.01 $& $0.10 \pm 0.02 $& $16.1 \pm
 0.2$\\
 \\
 Night 6 &Peak 3& $0.34 \pm 0.12 $& $0.09 \pm 0.03 $& $48 \pm 3$ \\
         &Peak 4& $1.07 \pm 0.12 $& $0.19 \pm 0.17 $& $15 \pm 2 $\\
         \\
Low Freq Avg & Peak 1\&3 & $ 0.34 \pm 0.01 $ &   & $ 49.0 \pm 0.1 $\\
High Freq Avg & Peak 2\&4 & $ 1.03 \pm 0.01 $ &   & $ 16.2 \pm 0.1 $\\
\hline
\end{tabular}
\end{minipage}
\end{table}

\subsection{Multi-Lorentzian Analysis}

\begin{table*}
\centering
 \begin{minipage}{\textwidth}
 \centering
 \caption{Best Fit Parameters for multiple Lorentzian Model. Average of Feature 1 considers Night 1 and Night 6. Average of Feature
 2 includes Night 2 and Night 6.}
 \centering
 \label{table:LorentParameter}
  \begin{tabular}{@{}c|c|cc|ccc@{}}
  \hline
  Number & White Noise & \multicolumn{2}{|c|}{VLF Noise}   &  \multicolumn{3}{|c|}{Feature 1} \\
  \hline
   & $\Gamma$ & $\nu_{max}$ & $r$ & $\nu_{0}$ & $\nu_{max}$ & $r$  \\
   & ($10^{-3}$) & mHz & ($10^{-3}$) & mHz & mHz & ($10^{-3}$) \\
  \hline
  Night 1 & 2.66$\pm{0.06}$ & 0.25$\pm{0.13}$ & 0.93$\pm{0.02}$ & 0.35$\pm{0.01}$ & 0.37$\pm{0.01}$ & 0.62$\pm{0.03}$ \\

  Night 2 & 4.84$\pm{0.08}$ & 0.02$\pm{0.04}$ & 0.12$\pm{0.23}$ & 0.32$\pm{0.08}$ & 0.46$\pm{0.03}$ & 0.41$\pm{0.03}$ \\

  Night 6 & 5.56$\pm{0.13}$ & 0.03$\pm{0.02}$ & 0.01$\pm{0.01}$ & 0.34$\pm{0.02}$ & 0.37$\pm{0.01}$ & 0.42$\pm{0.02}$  \\
  \hline
Average frequency&      &      &      & \multicolumn{3}{|l|}{0.35 $\pm 0.01$ \footnote{Average of night 1 and night 6 only.}} \\
\hline
 Average period &      &      &      & \multicolumn{3}{|l|}{48 $\pm 2$ min}\\
\hline

Number &  \multicolumn{3}{|c|}{Feature 2} &  \multicolumn{3}{|c|}{Feature 3} \\
\hline
&   $\nu_{0}$ & $\nu_{max}$ & $r$ & $\nu_{0}$ & $\nu_{max}$ & $r$ \\
 & mHz & mHz & ($10^{-3}$) & mHz & mHz & ($10^{-3}$)\\
\hline
Night 1& 1.17$\pm{0.01}$ & 1.18$\pm{0.01}$ & 0.10$\pm{0.03}$ & 3.37$\pm{0.15}$ & 3.64$\pm{0.15}$ & 0.09$\pm{0.02}$ \\
Night 2 & 1.03$\pm{0.01}$ & 1.03$\pm{0.01}$ & 0.43$\pm{0.06}$ & 2.94$\pm{0.10}$ & 3.46$\pm{0.04}$ & 0.10 $\pm{0.01}$ \\
Night 6& 0.99$\pm{0.03}$ & 1.01$\pm{0.01}$ & 0.11$\pm{0.01}$ & 2.64$\pm{0.50}$ & 3.76$\pm{0.20}$ & 0.14$\pm{0.01}$ \\

\hline

  Average frequency&
  \multicolumn{3}{|l|}{1.03 $\pm 0.01$ \footnote{Average of night 2 and night 6 only.}}&
  \multicolumn{3}{|l|}{3.46 $\pm 0.04$ \footnote{For broad features, $\nu_{max}$ is considered.}}\\
  \hline
    Average period &
  \multicolumn{3}{|l|}{16.2 $\pm 0.2$ min}&
  \multicolumn{3}{|l|}{4.8 $\pm 0.1$ min}\\
  \hline
    \end{tabular}

  \end{minipage}

\end{table*}

Where a mixture of periodic, quasi-periodic, and aperiodic variability
is present, as may be the case in these data, a superposition of
multiple Lorentzians has become widely used for power spectral
analysis in LMXBs
\citep{2000MNRAS.318..361N,2002ApJ...568..912V,2002ApJ...572..392B}.
While these techniques were developed to describe X-ray data, there is
no reason that they should not also be applicable to optical data too.  The
primary advantage of a Lorentzian description of the variability
components is that all quasi-periodic oscillations (QPOs) and noise
components can be fitted in the same way; it is not necessary to
decide the nature of each component in advance. Following
\citet{2000MNRAS.318..361N}, an individual Lorentzian component is
given by,
\begin{equation}\label{Eq:lorent}
    P(\nu) = \frac{r^{2}\Delta}{\pi}\frac{1}{\Delta^{2}+(\nu-\nu_{0})^{2}}
\end{equation}
where $r$ is the integrated fractional r.m.s. of each Lorentzian, and
$\Delta$ is its half width at the half maximum (HWHM).

\begin{figure}
\begin{center}
\begin{minipage}{\linewidth}
\centering
\epsfig{file=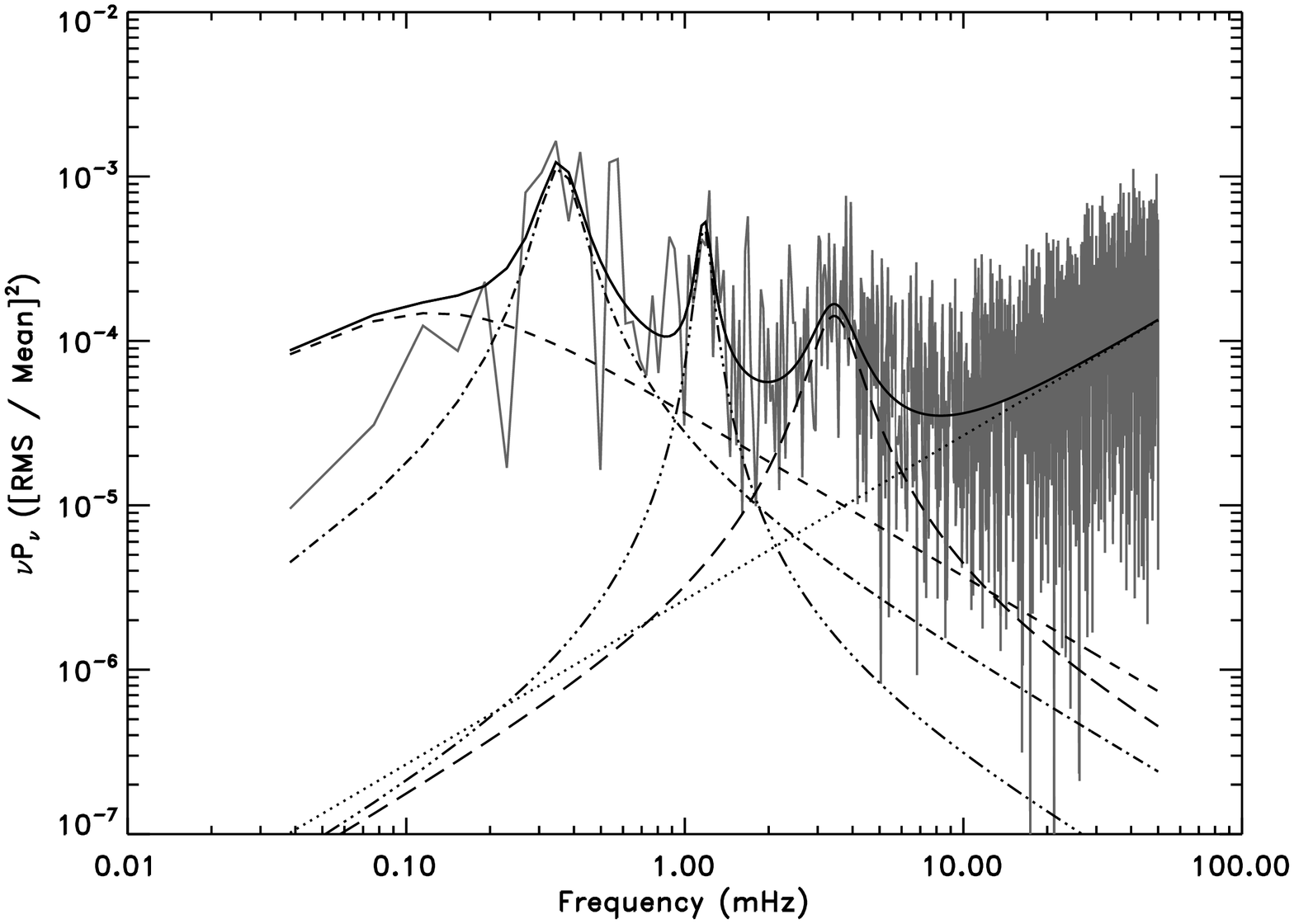,angle=90,width=\textwidth}
\epsfig{file=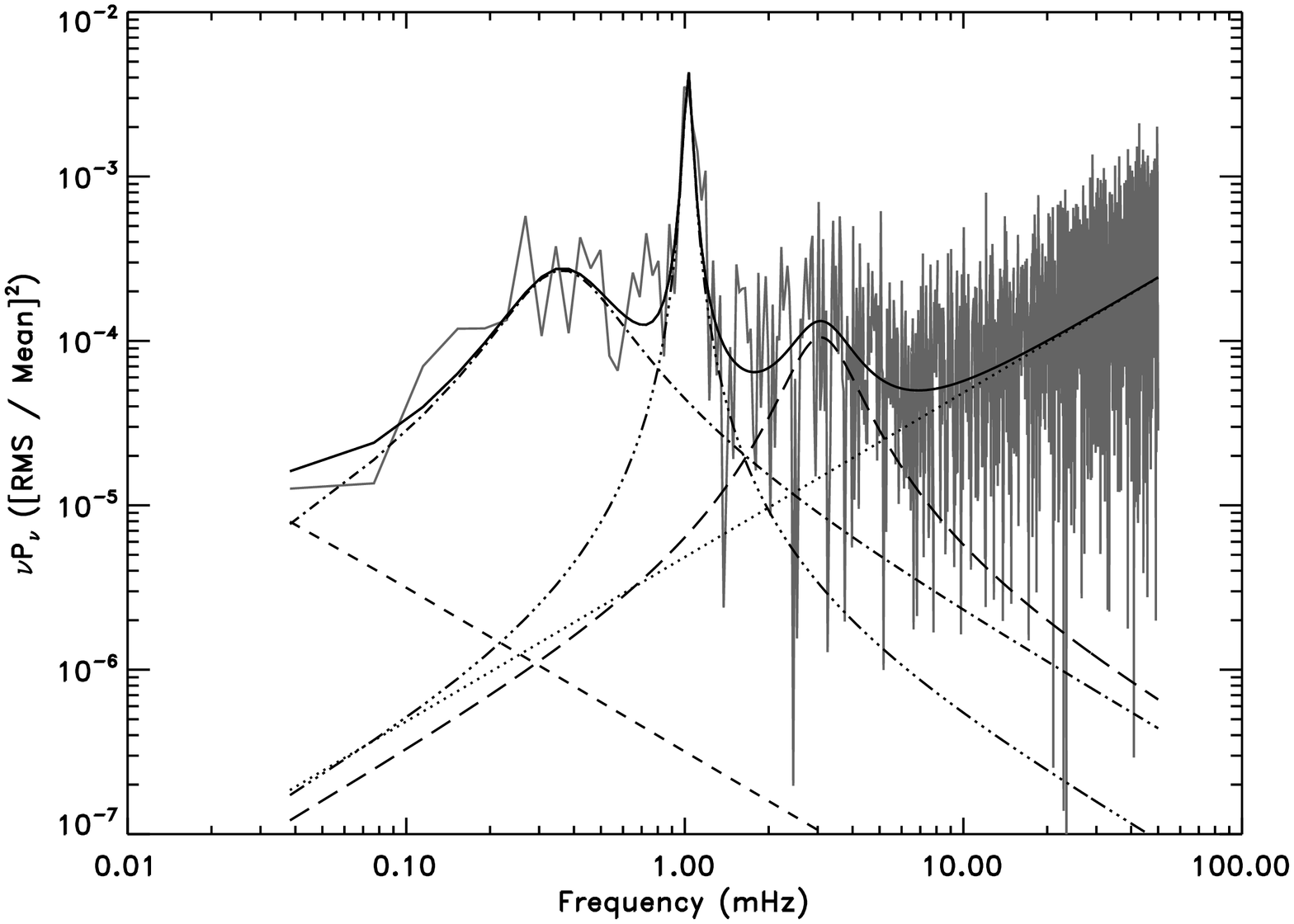,angle=90,width=\textwidth}
\epsfig{file=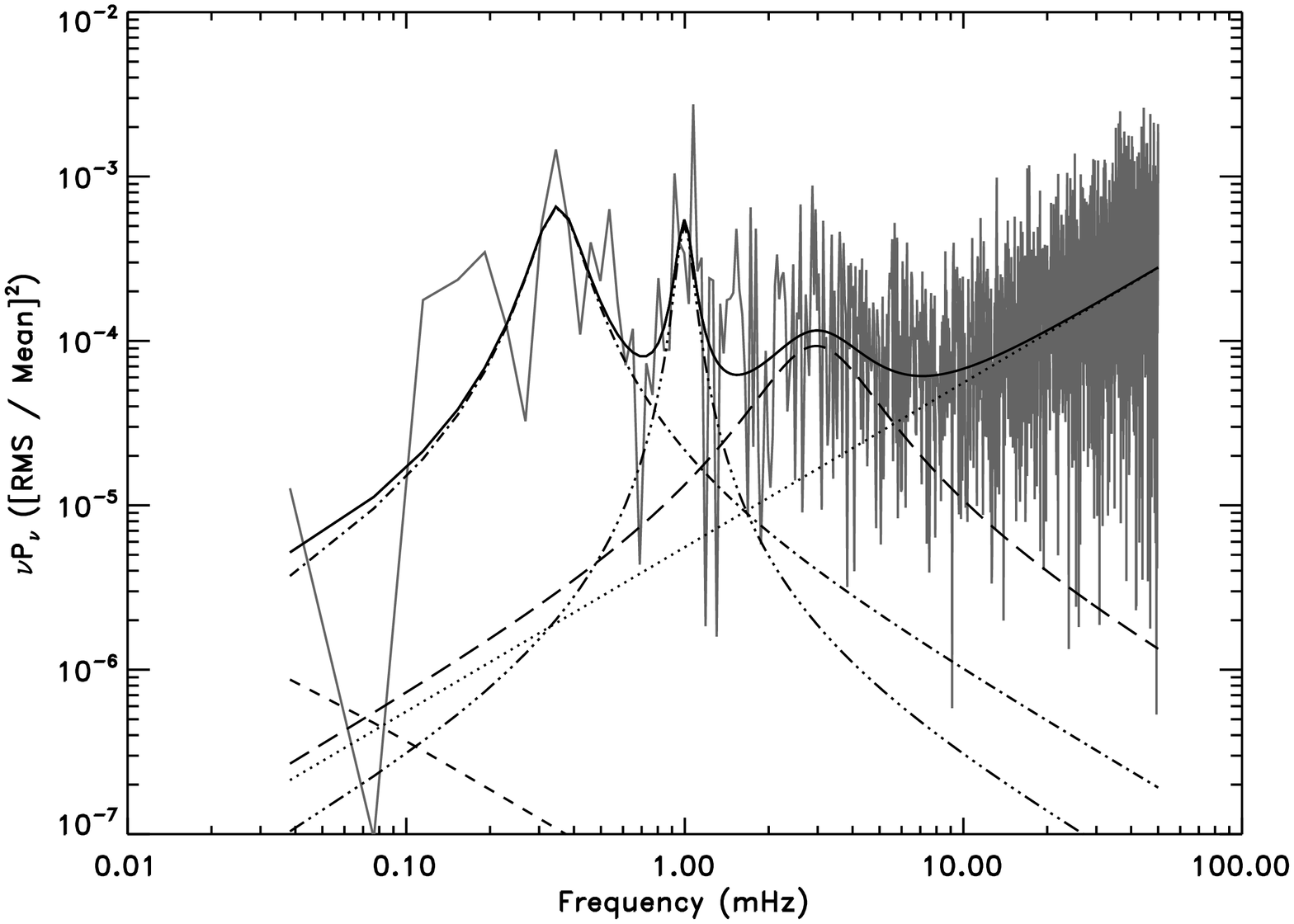,angle=90,width=\textwidth}
\end{minipage}
\end{center}
\caption{Power spectra and fit functions in the power spectral
density times frequency representation for three observations (top panel: night 1; middle panel: night 2; bottom panel: night 6). The
different dotted lines mark the individual Lorentzian components
of the fit. The dark line displays the final model.}
\label{fig:sp_log}
\end{figure}

We show in Figure~\ref{fig:sp_log} the power spectra and the fitting
functions in the ($\nu P_{\nu}$)
representation \citep{2000MNRAS.318..361N}, where the power spectral
density is multiplied by its Fourier frequency. It is helpful to
associate a characteristic frequency with each Lorentzian component.
We follow X-ray convention and use $\nu_{\rm max}$ as the characteristic
frequency for broad features.  This is the maximum in the $\nu P_{\nu}$
representation and defined by,
\begin{equation}\label{Eq:vmax}
    \nu_{\rm max} = \sqrt{\nu_{0}^{2}+\Delta^{2}}=\nu_{0}\sqrt{1+\frac{1}{4Q^{2}}}
\end{equation}
see \cite{2002ApJ...572..392B}, where $Q$ is the quality factor,
defined as $Q \equiv \nu_{0}/2\Delta$. For narrow features,
$\nu_{max} \approx \nu_{0}$.

Previous research using this technique on different sources was done in a higher
frequency range, 0.01--100\,Hz, typical of frequencies seen in X-ray data, e.g.
\cite{2002ApJ...568..912V}, \cite{2002ApJ...572..392B}. These authors have
found correlations between characteristic frequencies and break
frequencies of low and high frequency noise.
Our optical results show similar complexity, with multiple components
required, but in a lower frequency range than typically studied in
X-ray observations, exploiting the longer uninterrupted time-series
obtainable from the ground.

We found that to obtain a satisfactory fit,
four Lorentzian components plus
statistical white noise were needed.
We began with the two frequencies already isolated, but found that we
could not obtain a fit without excess power remaining at both higher
and lower frequencies. One zero-centered Lorentzian component was
added for very low frequency noise (VLF noise); this is a band-limited
noise component. The last Lorentzian component is fitting the higher
frequency peak revealed only in the $\nu P_{\nu}$ representation as
residual power around 3\,mHz. The characteristic frequencies and
integrated noise are listed in Table~\ref{table:LorentParameter}. The
quoted errors in $\nu_{0}$, $\nu_{max}$ and $r$ are 1\,$\sigma$
confidence limits.

As clearly seen in Figure~\ref{fig:sp_log}, the three peak features
can all be identified on each night in this representation. The first
two of them are relatively narrow features. Because no significant
differences are present between $\nu_{0}$ and $\nu_{max}$, we take
$\nu_{max}$ as the frequency of the feature. Feature 1 with an average of
0.35\,mHz is most significant on Night 1 and Night 6.  Hence, we
ignore results from Night 2 for this period. The second sharp feature
with average of 1.03\,mHz is especially strong on Night 2 but is also
present on Night 6. Results from Night 1 are ignored for Feature
2. Generally, the average results for these features are very consistent with
results from Gaussian fitting of the linearly scaled power spectra.
Feature 3 is quite broad. The average of $\nu_{max}$ from all three
nights is calculated. It is centered at 3.46\,mHz with about a 4\%
uncertainty.

\subsection{Phase stability of the 1\,mHz QPO}

\begin{figure}
\begin{center}
\begin{minipage}[!hb]{\linewidth}
\centering
\epsfig{file=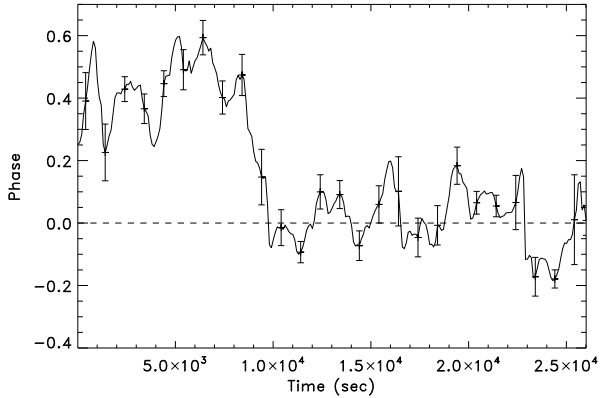,angle=90,width=\textwidth}
\end{minipage}
\end{center}
\caption{Phase Plot of Night 2 Period 1.03 mHz using 1000 sec as
window size and step is 100 sec. The error bars are measured for each 1000 sec window.} \label{fig:phase}
\end{figure}

Unlike the other features, the 1\,mHz QPO can be clearly seen in the
lightcurves, allowing direct study in the temporal domain.  In order
to better understand the 1\,mHz QPO, we examined the stability of its
phase. To do this we performed a sliding sine wave fit to the Night 2
data. We used a 1000\,sec window shifting with small step sizes of
100\,sec and fitted a sine wave fixed at the 16.2\,min period, but
with its phase and amplitude as a free parameter. The change in phase
with time is shown in Figure~\ref{fig:phase}. An unexpected phase change
occurs around 9500\,sec.  The phase is basically separated into two
segments, differing by 0.34 in phase. As a check, we also examined
individual segments of the light curve by eye and confirmed that the
phase change can be seen at this time.  This confirms that the signal
is quasi-periodic in nature and not a true period.

\section{Comparison with X-ray data}
\label{XraySection}

Our observations lie outside of the frequency range commonly studied
in X-rays.  For example, \citet{2002ApJ...568..912V} terminate the
power spectra densities (PSDs) of 4U~0614+091 around 10\,mHz.  Some of their PSDs do reveal a
very low frequency noise component overlapping with our frequency range.

We have searched for an X-ray mHz QPO with contemporaneous X-ray observations from the {\it Rossi X-Ray Timing Explorer} ({\it RXTE}). We chose the Proportional Counter Array (PCA) B band data, which is concentrated in a low energy band, 2--5\,KeV and examined data from dates nearest to our observation, from Dec 13--15 2007. Lightcurves were produced in the usual way using HEASoft version 6.8.  Lightcurves were extracted in 16\,sec bins, with a maximum duration of 51.2\,min.

\begin{figure}
\begin{center}
\begin{minipage}[h]{\linewidth}
\centering

\epsfig{file=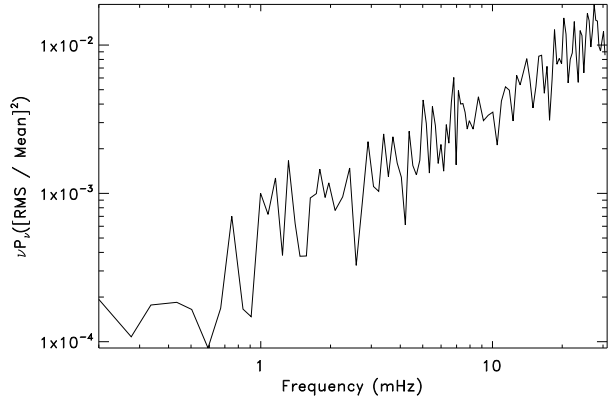,angle=90,width=\textwidth}

\end{minipage}
\end{center}
\caption{
Power-spectrum of RXTE PCA B band data from Dec 13--15 2007.
} \label{fig:pca}
\end{figure}

We used the lightcurves to generate low-frequency power-spectra, and
show these in Fig.~\ref{fig:pca}.
There is no evidence for an X-ray oscillation coincident with our
1\,mHz optical feature, although this may be a consequence of the
non-simultaneity of the data. This is similar to the situation in
4U~1626-67 \citep{2001ApJ...562..985C}. This is a 42\,min orbital
period system showing a strong 15 min UV and optical QPO without an
X-ray counterpart.

\section{Discussion}
\label{DiscussionSection}

Including the new results here, there are now four independent
observations supporting an optical periodicity of approximately 50\,min
(\citealt{OB05,2006MNRAS.370..255N,2008PASP..120..848S} and our
results).  Besides confirming the nature of 4U~0614+091 as an
ultracompact source, identification of the 50\,min period as orbital
would have important
implications for the nature of the companion star.  We should keep in
mind, however, that a periodicity that is detected repeatedly is not
necessarily an orbital period.  The possible presence of it in
spectroscopic data would suggest an orbital origin, but this detection
was very marginal \citep{2006MNRAS.370..255N}.

\citet{2004MNRAS.348L...7N} suggested that the companion was a CO
white dwarf based on optical detections of carbon and oxygen lines,
but an absence of hydrogen and helium.  \citet{2010MNRAS.401.1347N}
reconsidered this argument based on chemical abundance predictions
from new models and favour either a hybrid white dwarf or a very
evolved helium star donor.  For a 50\,min orbital period, these two
possibilities become observationally similar.

An additional consideration is that the presence of type I X-ray
bursts suggests the presence of helium in the accreted material
\citep{2010A&A...514A..65K}.  Both helium star and hybrid white dwarf
models could include mass-transfer phases where CO-enriched material
with a non-negligible helium fraction is transferred, although the
upper limits on the helium fraction implied by the non-detection of
helium lines seem in contradiction to the helium fraction required for
X-ray bursts \citep{2010A&A...514A..65K}.  The presence of apparent
helium bursts in helium deficient UCXBs appears to be a recurrent
problem \citep{2003ApJ...599..498J}. These authors proposed a generic solution to this problem in which CO rich material is accreted but then spallation reactions near the surface of the neutron star break these elements up in to hydrogen and helium which can then be burnt back to heavier elements during bursts \citep{1992ApJ...384..143B}.  We note that this mechanism will not work in pure CO material as spallation requires elements of quite different atomic number to be present.  If some helium is present in the accreted material, however, then this mechanism could potentially work to increase the helium fraction above that seen in the disk spectra.

A third possibility for a 50\,min binary would be that the donor star
is an evolved main-sequence star that has lost most of its hydrogen
envelope.  In this case, however, we would expect CNO processed
abundances with nitrogen more abundant than carbon and oxygen, in
contradiction to optical spectroscopy \citep{2004MNRAS.348L...7N}, so
this possibility can probably be discounted.

A major difficulty with identifying the donor as a white dwarf or
evolved helium star comes in the expected X-ray luminosity of the
system.  X-ray observations indicate an average luminosity of
$3.6\times10^{36}$\,erg\,s$^{-1}$.  In a neutron star system with no
event horizon the X-ray luminosity should be a reasonable indicator of
mass transfer rate, and in any case a lower-limit on it.  One
uncertainty remaining is that 4U~0614+091 does eject jets
\citep{2006ApJ...643L..41M}.  A more comprehensive study, however,
suggests that only 10\,percent of the accretion luminosity is being
carried by the jets \citep{2010ApJ...710..117M}.  This constraint, together
with the observed X-ray luminosity, implies a mass transfer rate of
$3\times 10^{-10}$\,M$_{\odot}$\,yr$^{-1}$.  This is much larger (by
an order of magnitude) than expected for the hybrid white dwarf or
evolved helium star models at 50\,min and evolving to longer periods
(\citealt{2010MNRAS.401.1347N}).  The inferred mass transfer rate
would be about right for an evolved main-sequence star, but as
discussed above this appears inconsistent with the chemical
abundances.

It is certainly possible that the 50\,min periodicity is not the
orbital period,
as none of the measurements demonstrate that the deduced modulations
have to be orbital in origin.  A white dwarf or helium star scenario would give
the right mass transfer rate for a shorter period below $\sim30$\,mins
(as earlier suggested by \citealt{2004MNRAS.348L...7N}).  In this
case, our mHz QPO would be a candidate for the orbital period, and
might then be attributed a superhump \citep{2001MNRAS.321..475H}.  It
is probably not the orbital period which should be strictly coherent,
as we have seen phase shifts.  Even in the case of a superhump,
however, it is hard to see how a phase shift such as this could arise
since it would involve a readjustment of the accretion disk on close
to its dynamical timescale.  If the orbital period is indeed below
30\,min, we then also have to explain the origin of the recurrent
50\,min timescale.

A more natural explanation for a large amplitude mHz oscillation seen
in the optical but not X-rays, and that is able to undergo large phase
shifts in a single cycle is that it originates further into the
accretion disk but
in regions too cool to produce X-rays.  Optical light in 4U~0614+091
appears to be dominated by the accretion disk
\citep{2006ApJ...643L..41M,2010ApJ...710..117M} so participation by the
disk is probably necessary to obtain a 10\,percent amplitude.  The
frequency and strong amplitude of our mHz oscillation strikingly
resembles that seen in 4U~1626--67 \citep{2001ApJ...562..985C}.  That
system too shows a CO spectrum and is identified as likely containing
a hybrid white dwarf or helium star donor \citep{2010MNRAS.401.1347N}.
The orbital period of 42\,min is a close match to that proposed for
4U~0614+091, but much more secure since it is seen in pulse-timing
residuals.  For this orbital period the mass transfer rate inferred
for a white dwarf donor is also about an order of magnitude greater
than that expected from gravitational radiation
\citet{1998ApJ...492..342C}, a problem similar to that encountered if
we attribute a 50\,min orbital period to 4U~0614+091.

\citet{2001ApJ...562..985C} suggested that the oscillation in
4U~1626--67 arises from a purely geometric effect, the tilt angle of
the normal to the local disk surface varies with azimuth. The
reprocessed emission would be expected to be sensitive to warping
material along the line of sight. \citet{2002ApJ...565.1134S}
suggested the magnetic torque from the pulsar interacting with the
accretion disk may produce the warping effect.  The similarity of
behavior in 4U~0614+091, however, appears at odds with this as there
is no evidence that this system also harbors a neutron star with a
significant magnetic field.  A possible solution is proposed by
\citet{2003ApJ...591L.119L}.  In the absence of a magnetized compact
object, it is suggested that magnetic fields threading the disk itself
could cause magnetically driven warping.  This idea was proposed in
the context of systems with a magnetically-driven outflow. Multiwavelength observations suggest the presence of a jet-like outflow is 4U~0614+091 \citep{2006ApJ...643L..41M,2010ApJ...710..117M} and so this may be a natural system in which to invoke this effect.

As a final remark, we note that the mHz oscillation is not seen alone,
but as one of three frequencies that are recurring from night to
night.  While one of these may be orbital, it is possible that none
are.  We are thus not looking for a mechanism that can give rise to a
single oscillation, but to a more complex multi-periodic behavior,
rather reminiscent of that see at much higher frequencies in X-rays
\citep{2002ApJ...568..912V}.  This would suggest a more complex
multi-modal warping of the accretion disk as a possible mechanism.  We
have focused on the 1\,mHz QPO here because it can be seen clearly in
the lightcurves, and so studied in the time-domain.  The 3.5\,mHz QPO
is certainly also of interest but with less information, and no
obvious timescale to associate it with, it is harder to say anything
about it.  More observations will be needed to properly disentangle
the behavior seen, for example by looking for correlations between
changes in the oscillation frequencies to test if they are linked or independent.

\section{Conclusions}

We have identified three persistent, distinct, periodic or
quasi-periodic oscillations in the optical flux from 4U~0614+091. The
0.35 mHz (48 min) period was suggested to be an orbital period by
\citet{OB05} and \citet{2008PASP..120..848S}, and supported by
\citet{2006MNRAS.370..255N}. If it is indeed orbital, it would confirm
4U~0614+091 as an ultracompact binary,
\citep{2001ApJ...560L..59J,2004MNRAS.348L...7N}.

The origin of the 1\,mHz QPOs now seen in both 4U~0614+091 and
4U~1626--67 remains unknown, if they are even related to each other.
It likely originates in the accretion disk and could arise as the
precession period of inner disk regions that are not hot enough to
produce X-ray emission.  The less prominent 3\,mHz QPO may be related
to this, or may be distinct.

\section{Acknowledgments}

Support for these observations was provided by a Louisiana State
University Council on Research Faculty Research Grant.  R.I.H. also
acknowledges support from NASA/Louisiana Board of Regents grant
NNX07AT62A/LEQSF(2007-10) Phase3-02 and National Science Foundation
Grant No.\ AST-0908789.  This research has made use of data obtained
through the High Energy Astrophysics Science Archive Research Center
Online Service, provided by the NASA/Goddard Space Flight Center and
has also made use of the NASA ADS Abstract Service.

\label{lastpage}

\end{document}